\documentclass[12pt]{article}
\usepackage[T1]{fontenc}
\usepackage[utf8]{inputenc}

\usepackage[backend=biber,
  sorting=none, 
  autocite=plain, 
  url=true,                  
  hyperref=true,
  style=numeric-comp]{biblatex}
\usepackage{graphicx}
\usepackage{subfig}
\usepackage{amsmath}
\usepackage{amssymb}
\usepackage[section]{placeins}
\usepackage{pgfplots}
\pgfplotsset{compat=1.12}
\usepackage{xcolor}
\definecolor{ballblue}{rgb}{0.13, 0.67, 0.8}

\usepackage{bm}
\usepackage{physics}

\usepackage{mathtools}
\usepackage{esvect}

\usepackage{fontawesome}
\usepackage{enumitem}

\usepackage[margin=0.75in]{geometry}

\usepackage[colorlinks=true, citecolor=blue]{hyperref}

\usepackage{authblk}
\usepackage{booktabs}
\usepackage{caption}
\title{{\bf The Aharonov-Bohm effect for a constant scalar matter potential in neutrino flavour interferometry}}

\author[1]{Jos\'e Bernab\'eu\footnote{Jose.Bernabeu@uv.es}}
\author[2]{Catalina Espinoza\footnote{m.catalina@fisica.unam.mx}}
\affil[1]{Department of Theoretical Physics, University of Valencia, and IFIC, Joint Centre UV-CSIC,
		Burjassot,  Val\`encia, E-46100, Spain.}
\affil[2]{C\'atedras Conahcyt - Department of Theoretical Physics, Instituto de F\'{i}sica, UNAM, Apdo. Postal 20-364, 01000 CDMX, M\'{e}xico.}
\date{}                   
\newcommand{\hi}[1]{\bf{#1}}

\begin{document}
	
\maketitle
	
\begin{abstract}

The Aharonov-Bohm effect is one of the most surprising wonders
of the quantum world. The observed solenoid effect, as well as others,
shows that a particle is affected by the potential in a region in which there is no force-field. This is so through the phase of the probability amplitude. Its interpretation is debated between a physical significance of the potential versus non-locality of quantum physics with the presence of the force-field generated by the potential difference outside this region. We demonstrate that the debate is resolved with the idea of replacing spatial interference by flavour interferometry as observed in neutrino  oscillations. The neutrino propagation through the crust of the Earth
in current facilities is affected by a constant scalar matter potential
and the phase difference is in flavour internal properties. Here we show how to signal and experimentally disentangle the phase-shift due to the potential by means of observables characteristic of its symmetry properties. The energy dependence of the neutrino-antineutrino asymmetry
in the golden transition $\nu_\mu \rightarrow \nu_e$ allows a clear separation of the matter component against the genuine asymmetry in the absence of the potential. These findings define, in a perfect symbiosis between quantum physics and particle physics, the path to observe in a single experiment the physical significance of the potential, the neutrino mass hierarchy and the genuine matter-antimatter asymmetry in the lepton sector.

\end{abstract}

\section{Introduction}
\label{intro}

The Ehrenberg-Siday-Aharonov-Bohm effect \autocite{W_Ehrenberg_1949,Aharonov:1959fk}, in which a charged particle is affected by an electromagnetic potential $(V, {\mathbf{A}})$ in a region with neither a magnetic field ${\mathbf{B}}$ nor an electric field ${\mathbf{E}}$, is one of the Seven Wonders of the  Quantum World \autocite{brooks2010seven}. The underlying mechanism is in the phase of the wave function, so the Aharonov-Bohm effect can be observed by interference experiments. The most commonly described case, the Aharonov-Bohm solenoid effect, takes place when the wave function of a charged particle passing around a long solenoid experiences a phase shift as a result of the enclosed magnetic field. This is so despite the magnetic field being absent in the region through which the particle travels. This phase shift has been observed \autocite{Chambers:1960xlk,mollenstedt1962messung}. An electric Aharonov-Bohm phenomenon was also predicted \autocite{batelaan2009aharonov}, in which a charged particle is affected by regions with different scalar potentials and zero electric field, but this has no experimental confirmation yet. Another experiment involving a ring geometry interrupted by tunnel barriers, with a bias voltage $V$ relating the potentials of the two halves of the ring, leads to an Aharonov-Bohm phase shift and it has been observed \autocite{van1998magneto}.  Recently \autocite{Overstreet:2021hea}  the split of a cloud of cold atoms into two atomic wave packets, one of them subjected to gravitational interaction with a large mass, has been reported with an observed phase shift due to a gravitational Aharonov-Bohm effect.

The prevailing interpretation of the effect is the physical significance of the potentials in quantum mechanics governing the phase of the probability amplitudes given by the Action. However, in the solenoid effect for example, the result can be written in terms of the enclosed magnetic field, leading to a school of thought giving the weight of the argument to non-locality of quantum mechanics. The debate is provoked by the formalism itself via the Stoke's theorem, because the phase shift - the induced relative phase between the two interfering amplitudes - is given by

\begin{equation}\label{eq1}
	\frac{q}{\hbar} \oint {\mathbf{A}}\cdot d{\mathbf{l}} = \frac{q}{\hbar} \Phi_B,       
\end{equation}
where $q$ is the particle charge and $\Phi_B$ the enclosed magnetic flux.                               
The alternative is thus centered between a local effect of the potential  - the left-hand side of Eq. (\ref{eq1}) - against a non-local effect of the force field - the right-hand side of Eq. (\ref{eq1}) -.

In this paper we clarify and close this debate with the idea of
replacing spatial interferometry by the flavour interferometry as observed in neutrino oscillations. In quantum physics {\hi{the interference is between probability amplitudes for different untagged alternatives, no matter whether the observables refer to either spatial or internal properties.}} Thus we consider a type of Aharonov-Bohm effect in which the phase difference between the ``arms of the interferometer'', generated by a {\hi{constant}} scalar potential, is in no way associated to the existence of any near or far force-field. This is made by using an appropriate interpretation \autocite{Bernabeu:1999ct,Banuls:2001zn} of the matter effect \autocite{Wolfenstein:1977ue,Mikheyev:1985zog} in neutrino propagation. In this propagation along ordinary matter with constant electron density, electron-neutrinos feel a constant scalar potential which is absent for the other muon and tau neutrino flavours. The different potential is generated by the charged-current weak interaction of electron-neutrinos with matter electrons. We demonstrate below that, in analogy with the Aharonov-Bohm effect, a source of definite flavour neutrinos is addressed to an ``interferometer'' as given by the PMNS mixing matrix \autocite{Pontecorvo:1957cp,Pontecorvo:1957qd,Maki:1962mu} of neutrino flavours, connecting to ``slits'' of effective mass-eigenstates, leading to ``arms'' with definite quantum coherent propagation at  {\hi{different constant}} potentials. The arms interfere in  a ``screen'' represented by the detection of a given neutrino-flavour. The ideal aim is to identify experimentally the relative phase-shift induced by the matter potential, separating it from the intrinsic phase difference associated to neutrino oscillations in vacuum, due to the oscillation frequencies corresponding to different neutrino masses. We thus consider appropriate observables able to disentangle the effect of the matter potential from the free propagation. Needless to say, our ``arms'' all have the same spatial direction. There is no room to blame the effect as due to potential gradients in space. It is due to potential differences in flavours, with no intervention of a force-field.

In fact the matter potential in the Sun, different for electron-neutrinos and muon-neutrinos, has been already needed to explain the solar neutrino oscillations which, in a good approximation, involve these two flavours. However, neither a separate phase-shift effect is obtained nor the constant potential requirement is present. A direct search for matter effects in the Earth \autocite{Super-Kamiokande:2013mie} results in a $2.7$ $\sigma$ indication that the electron flavor content of solar neutrinos during nighttime, where Earth matter effects are present, exceeds that of daytime. But again one can argue that the matter potential is not constant when neutrinos traverse the core of the Earth. The condition of a constant potential is satisfied in terrestrial accelerator experiments with neutrino propagation in the crust of the Earth. Is it possible to separate the effect of the phase-shift due to the matter potential in these experiments?  We propose here observables which are characteristic of the matter potential, not accessible by interference effects in vacuum: {\hi{the CPT-odd component of discrete asymmetries in neutrino flavour oscillations.}}

\section{Symmetry properties disentangle the potential phase-shift by interference}

At present all parameters describing neutrino oscillations are well established, except the genuine CPV phase $\delta$ in the $U_{\textrm{PMNS}}$ mixing matrix. Mixing angles and neutrino mass differences are known \autocite{Esteban:2024eli} up to the sign of the neutrino mass hierarchy. In the basis of the three neutrino flavours, the Hamiltonian matrix is written  as

\begin{equation}\label{eq2}
	H = \frac{1}{2E} \left\{ U \left[ 
	\begin{array}{ccc}
		m_1^2 & 0 & 0 \\
		0 & m_2^2 & 0 \\
		0 & 0 & m_3^2
	\end{array}
	\right] 
	U^\dagger +\left[ 
	\begin{array}{ccc}
		a & 0 & 0 \\
		0 & 0 & 0 \\
		0 & 0 & 0
	\end{array}
	\right] 
	\right\}	
	= \frac{1}{2E} \tilde{U} \tilde{M}^2 \tilde{U}^\dagger,
\end{equation}
where $a = 2E V$, with $E$ the relativistic neutrino energy and $V$ the interaction
potential having a different effect in the three effective neutrino masses for the ``arms'' of the propagation in the last side of Eq. (\ref{eq2}). $\tilde{U}$ and $\tilde{M}^2$ refer to effective values in
matter. For antineutrinos, $a$ changes sign and $U$ changes to its conjugate. Due to the energy mismatch between the two terms of $H$, all effective values become energy dependent. The effective values of neutrino masses in matter and those of the rephasing-invariant mixings in matter can be connected with the parameters in vacuum plus the matter potential  present in the Hamiltonian of Eq. (\ref{eq2}). Actual experiments, like DUNE \autocite{DUNE:2020jqi} or Hyper-Kamiokande \autocite{Hyper-Kamiokande:2016srs}, cover neutrino energies in a region such that a hierarchy exists between $a$ and the neutrino mass differences

\begin{equation}\label{eq3}
	\Delta m^2_{21} < |a| < | \Delta m^2_{31} |.
\end{equation}

The corresponding hierarchy for the induced phases  $\Delta_{21} = \frac{\Delta m^2_{21} L}{4E}$,
$A = \frac{aL}{4E}$ and $\Delta \equiv \Delta_{31} = \frac{\Delta m^2_{31} L}{4E}$  suggests a perturbative expansion of the different transition probabilities with respect to the contribution generated by the $\Delta$ phase. Relative to it, we keep linear terms in $\Delta_{21}$, needed for bringing the CPV phase $\delta$ of the mixing matrix into operation, accompanied by the potential phase $A$ up to second order, i.e., $A$, $\Delta_{21}$, $A^2$ and $A \times \Delta_{21}$. This expansion, neglecting $A^3$, $A^2\times \Delta_{21}$ and higher order terms, allows a clear understanding of their contributions to the interesting asymmetries to be selected. Once constructed, the analytical perturbative result will be compared with numerical exact results to prove their good agreement. Taking into account that the potential in matter is odd under CP and CPT, changing sign for neutrinos and antineutrinos, we show in Table \ref{tab1} the definite symmetry properties of these terms
with T-even behaviour in the direct diagonal terms and T-odd behaviour
in the inverse diagonal terms. The reference term $1$ in Table \ref{tab1} corresponds to the contribution with the $\Delta$ phase in neutrino oscillations. Each of the terms, in the considered neutrino flavour transition, will be accompanied by the corresponding rephasing invariant mixing and a function $f(\Delta)$. The different terms linear in the potential phase-shift $A$, being CPT-odd, will have a common $\Delta$-dependent factor. As the T-even  asymmetry has to be even in the baseline $L$ whereas the T-odd asymmetry has to be odd in the baseline $L$, a corolary of Table \ref{tab1} tells us the parity of $f(\Delta)$ under the change of sign of $\Delta$: an odd function $f(\Delta)$ has a precious interest in order to fix the pending question of the hierarchy in the neutrino masses. We observe that {\hi{the first order term, the phase $A$ induced by the matter potential, contains an odd function of $\Delta$, so its sign fixes the neutrino mass hierarchy.}} This conclusion reinforces the interest of the present study to find observables able to signal the term with the phase $A$ of the matter potential. Notice that, due to the symmetry behaviour, there is no $A^2$ correction term in the asymmetries, thus expecting an excellent agreement with the exact numerical result.

\begin{table}[t]
	\caption{The symmetry behaviour of the different terms
		in the expansion of transition probabilities.}\label{tab1}%
	\centering
	\begin{tabular}{@{}c|c|c@{}}
		\toprule
		& CPT-even  &  CPT-odd \\
		\midrule
		CP-even    & $1$, $\Delta_{21} \times\cos(\delta)$, $A^2$   & $A\times\Delta_{21} \times\sin(\delta)$  \\
		\midrule
		CP-odd    &  $\Delta_{21} \times\sin(\delta)$   & $A$, $A\times\Delta_{21} \times\cos(\delta)$  \\
		\bottomrule
	\end{tabular}
\end{table}

A priori, we may consider three possible observables with a $\mu$-neutrino beam as reference. We give detailed expected results for a baseline of $L=1300$ Km., covering energies along the first and second oscillation peaks.

\section{The CPT-odd Asymmetry $P(\nu_\mu \rightarrow \nu_e) - P(\bar{\nu}_e \rightarrow \bar{\nu}_\mu)$}

This choice has the beauty that a non-vanishing observation of the asymmetry is, by itself, a necessary and sufficient condition  to
demonstrate the effect of the $A$ phase. Its experimental search
would need a neutrino factory facility with a charge discrimination detector. We discuss the expected results for this asymmetry, linearly proportional to the potential, as well as its separation into the independent and dependent terms of the CPV $\delta$ phase in the mixing matrix. We calculate % in {Methods 1} 
this asymmetry expanding cross-checking results \autocite{Banuls:2001zn,Akhmedov:2004ny,Bernabeu:2019npc} for the relevant transitions. We obtain

\begin{equation}\label{eq4}
	A_p^{(CPT)}(E) = A_1(E) + C_1(E) \cos(\delta) + S_1(E) \sin(\delta),     
\end{equation}
with the first two terms, being T-even and CP-odd, given by

\begin{equation}\label{eq5}
	A_1(E) = 16 \, A \,  S \,  [\sin(\Delta)/\Delta - \cos(\Delta)] \sin(\Delta),
\end{equation}

\begin{equation}\label{eq5b}
	C_1(E) = 16 \,  A \,  \Delta_{21} \,  J_r \,  [\sin(\Delta)/\Delta -\cos(\Delta)] \cos(\Delta),
\end{equation}
and the last term, being CP-even and T-odd, given by

\begin{equation}\label{eq6}
	S_1(E) =  - 16 \,  A \,  \Delta_{21} \,  J_r \,  [\sin(\Delta)/\Delta -\cos(\Delta)] \sin(\Delta),
\end{equation}
with the rephasing-invariant mixings $S=c^2_{13} \, s^2_{13} \, s^2_{23}$ 
and $J_r=c^2_{13} \, s_{13} \, c_{12} \, s_{12} \, c_{23} \, s_{23}$.

\begin{figure}[t]
	\centering
	\includegraphics[width=0.5\textwidth]{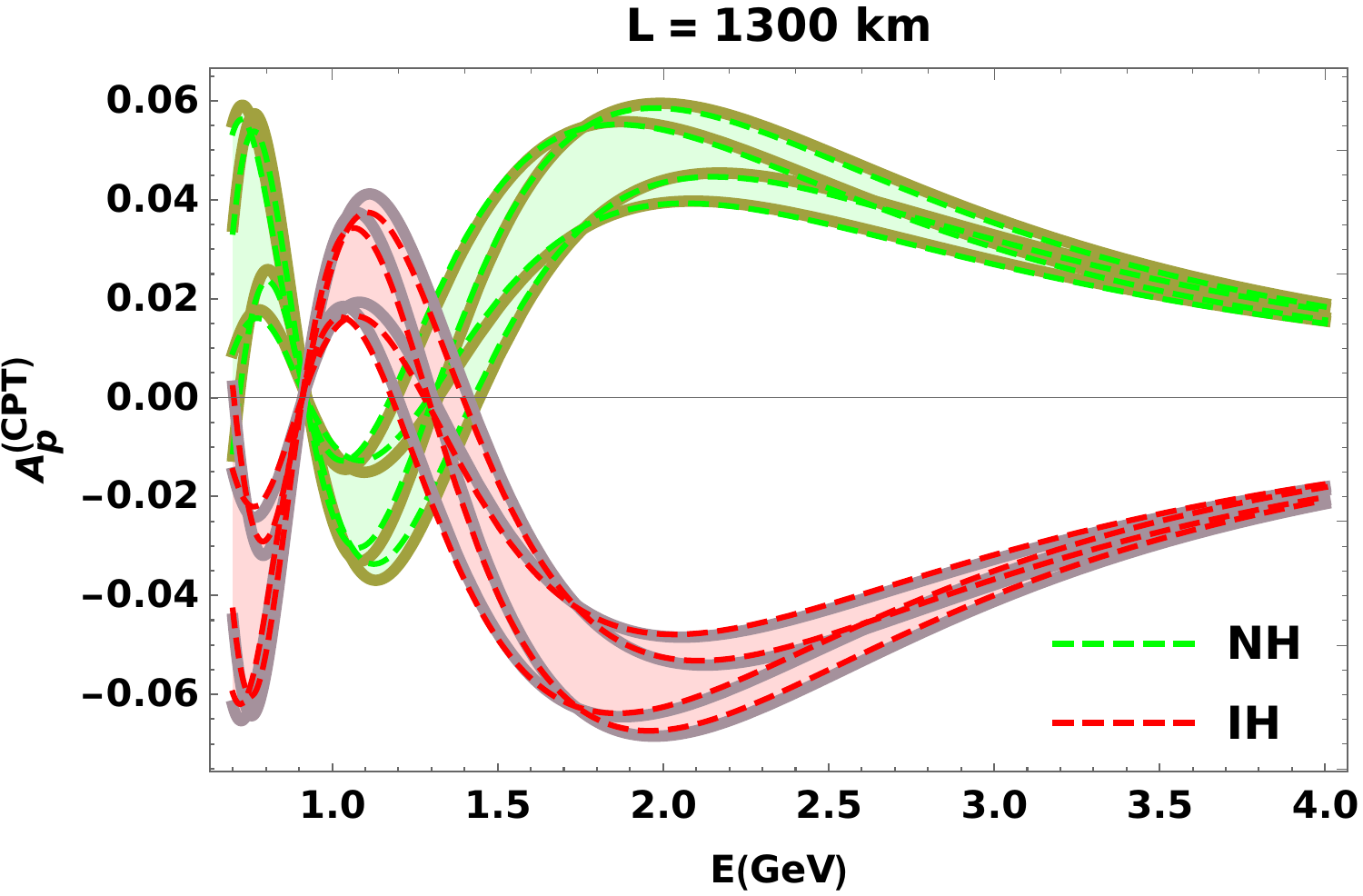}%
	\includegraphics[width=0.5\textwidth]{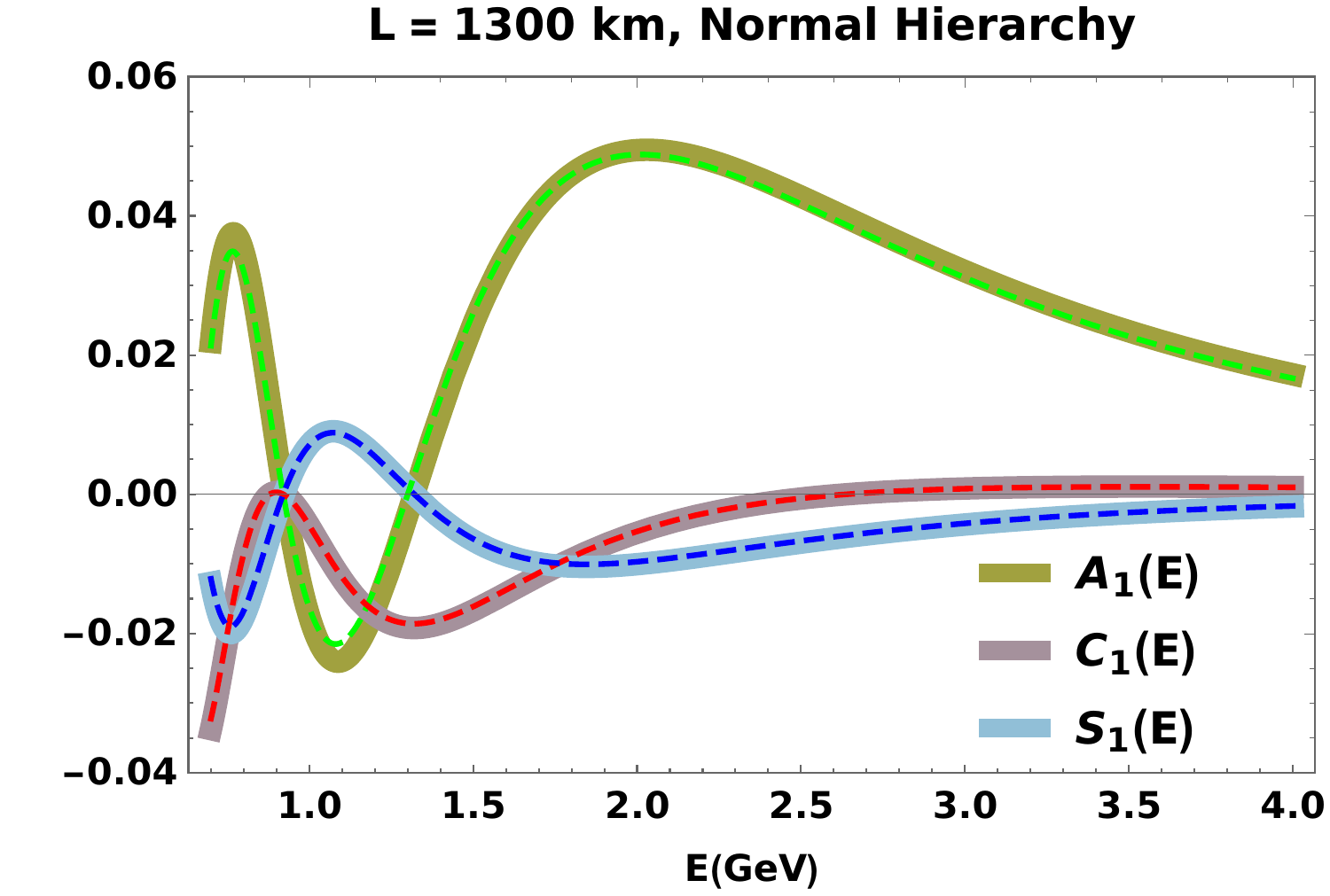}
	\caption{Left panel: The $A_p^{(CPT)}(E)$ asymmetry for normal - green -     
		and inverted - red - hierarchies.
		Right panel: The separate $A_1(E)$, $C_1(E)$, and $S_1(E)$ terms.
		Solid (dashed) lines portrait the approximate (numerical exact) equations.
		Colored bands depict the whole range of values of $\delta$.
	}\label{fig1}
\end{figure}

The observable $A_p^{(CPT)}(E)$ asymmetry is plotted, as function of energy $E$, in the left panel of Fig. \ref{fig1} for normal and inverted hierarchies, compared with the exact results (dashed lines). The green  - normal - and red - inverted -  bands include the results for all values of $\delta$, indicating the dominance of the $A_1(E)$ term in the entire energy region of the first oscillation peak.  The separate three terms $A_1(E)$, $C_1(E)$, and $S_1(E)$ are plotted in the right panel of Fig. \ref{fig1} confirming the preeminence of $A_1(E)$.

We remind the reader that $A_1(E)$ and $S_1(E)$ are odd in changing the sign of $\Delta$ whereas $C_1(E)$ is even under the change of sign of $\Delta$.

This CPT-asymmetry, proportional to the phase-shift $A$ induced by the
potential, has a peculiar energy-dependence not given by the canonical
oscillation peaks and it leads to a {\hi{magic energy}} \autocite{Bernabeu:2019npc} around the
solution of $\tan(\Delta) = \Delta$ where the asymmetry vanishes independent of $\delta$. For
$L = 1300$ Km. the magic energy is $0.92$ GeV, near the second oscillation peak of the transition probabilities. Needles to say, the proportionality of this asymmetry to $A$ and the preeminence of the $A_1$ term make it ideal for the aim of separating out the effect of the potential. However, there is no facility at present able to measure it.

\section{ The CPT-odd and CP-odd Asymmetry \scalebox{0.8}{%
		 $P(\nu_\mu \rightarrow \nu_\mu) - P( \bar{\nu}_\mu \rightarrow \bar{\nu}_\mu)$}}

\begin{figure}[t]
	\centering
	\includegraphics[width=0.5\textwidth]{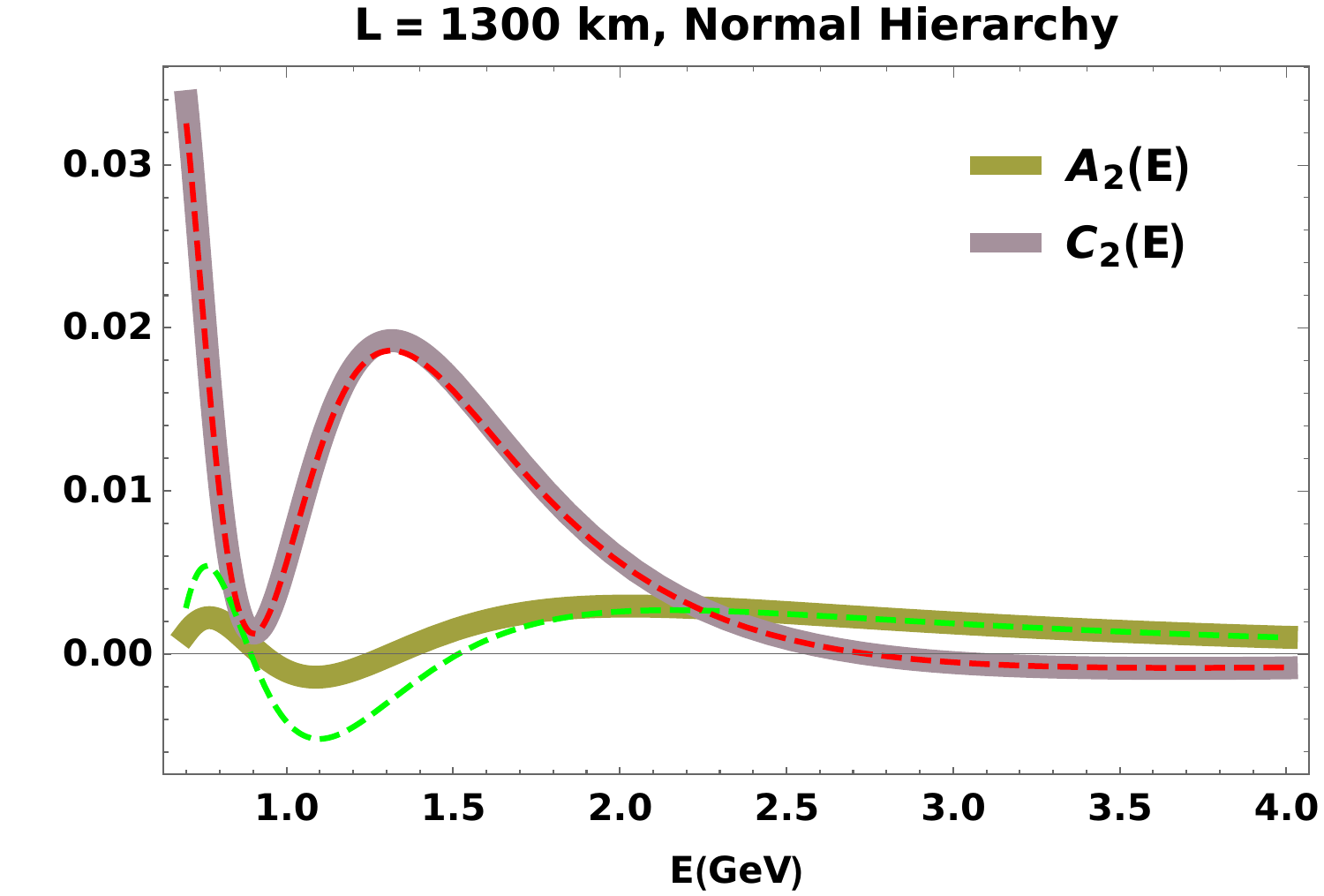}%
	\caption{The separate terms $A_2(E)$  - green - and $C_2(E)$ - red -
		of the $A_p^{(CPT, CP)}(E)$ asymmetry for normal
		hierarchy.
		Solid (dashed) lines portrait the approximate (numerical exact) equations.}\label{fig2}
\end{figure}

Again this alternative asymmetry, if observed, would represent a clear-cut confirmation of the effect of the potential in neutrino flavour interferometry. 
Although the effect of the potential in this disappearance experiment is expected to be small, we quantify here this statement, in view of being an indisputable effect of the potential for an observable in the DUNE experiment. 
We proceed to the discussed perturbative expansion of the survival probabilities \autocite{Banuls:2001zn,Akhmedov:2004ny}, leading to an asymmetry generated by the matter potential

\begin{equation}\label{eq7}
	A_p^{(CPT, CP)}(E) = A_2(E) + C_2(E) \cos(\delta),
\end{equation}
with both $A_2(E)$ and $C_2(E)$ terms being T-even. At the required perturbative order we get

\begin{equation}\label{eq8}
	A_2(E) =  16 A \, S \cos(2\theta_{23}) \left[ \sin(\Delta)/\Delta - \cos(\Delta) \right] \sin(\Delta)
\end{equation}

\begin{align}
	C_2(E) =& 16 A \, J_r \, \Delta_{21} \Bigl[
	\cos(2\theta_{23}) \sin^2(\Delta) 
	+ \left( \cos^2(\Delta) - \sin^2(\Delta)/\Delta^2 \right) \nonumber \\
	&+  2 s^2_{23} \left[ \sin(\Delta)/\Delta - \cos(\Delta) \right] \sin(\Delta)/\Delta
	\Bigr] \label{eq8b}
\end{align}
a result which explains its small effect due to a (2, 3) mixing around the maximum. An additional consequence is that its dominance by the
$C_2(E)$ term, which is an even function of $\Delta$, implies that this observable is not sensitive to the neutrino mass hierarchy, in spite of
its proportionality to the potential phase-shift. Its actual value is fundamentally affected by the CPV phase $\delta$ of the mixing matrix.
$C_2(E)$ is plotted in Fig. \ref{fig2} together with the expected $A_2(E)$ for normal
hierarchy.

As expected, the same magic energy appears  here. Although
this asymmetry in the disappearance experiment is proportional to $A$,
the small value of the $A_2(E)$ term in the entire energy region  advises against its use for the purpose of this paper.

\section{The CP-odd Asymmetry $P(\nu_\mu \rightarrow \nu_e) - P( \bar{\nu}_\mu \rightarrow \bar{\nu}_e)$}

These probabilities are those of the golden transitions, available in the current experiments on neutrino appearance oscillations. Contrary to the previous asymmetries, this entire CP-odd asymmetry is not proportional to the phase-shift $A$ induced by the matter potential. It is in addition induced by a genuine CP-violation in free neutrino oscillations, the main focus in DUNE and HK experiments. In spite of this combined effect at a given energy $E$ - and fixed baseline $L$ -, a recent study \autocite{Bernabeu:2018twl,Bernabeu:2018use}
%\autocite{Bernabeu:2019npc,Bernabeu:2018twl}  
has demonstrated a disentanglement theorem for the two components by means   of their different behaviour under the other discrete symmetries. The result for
$A^{(CP)}(E) = A_p^{(CP, CPT)}(E) + A_g^{(CP, T)}(E)$ is a well defined unique separation with $A_p^{(CP, CPT)}(E)$ being a CPT-odd and T-even component, whereas $A_g^{(CP, T)}(E)$ is a T-odd and CPT-even component. The separation is not only a theoretical construct, it is experimentally accessible because - even for a fixed baseline $L$ - {\hi{the two components show different energy dependence.}} 
We exploit these findings for giving the analytic perturbative results for the two components

\begin{equation}\label{eq9}
	A_p^{(CP, CPT)}(E) = A_3(E) + C_3(E) \cos(\delta),
\end{equation}

\begin{equation}\label{eq9b}
	A_g^{(CP, T)}(E) = S_3(E) \sin(\delta),
\end{equation}
with the two terms for the CPT-odd and T-even component

\begin{equation}\label{eq10}
	A_3(E) = 16 A \, S [\sin(\Delta)/\Delta-\cos(\Delta)] \sin(\Delta)
\end{equation}

\begin{equation}\label{eq10b}
	C_3(E) = 16 A \, \Delta_{21} J_r [\sin(\Delta)/\Delta - \cos(\Delta)] \cos(\Delta)
\end{equation}
and the result for $S_3(E)$ identical, up to the required order, to
the CPV-asymmetry for oscillation in vacuum

\begin{equation}\label{eq11}
	S_3(E) =  - 16 \Delta_{21} J_r \sin^2(\Delta)
\end{equation}

\begin{figure}[t]
	\centering
	\includegraphics[width=0.5\textwidth]{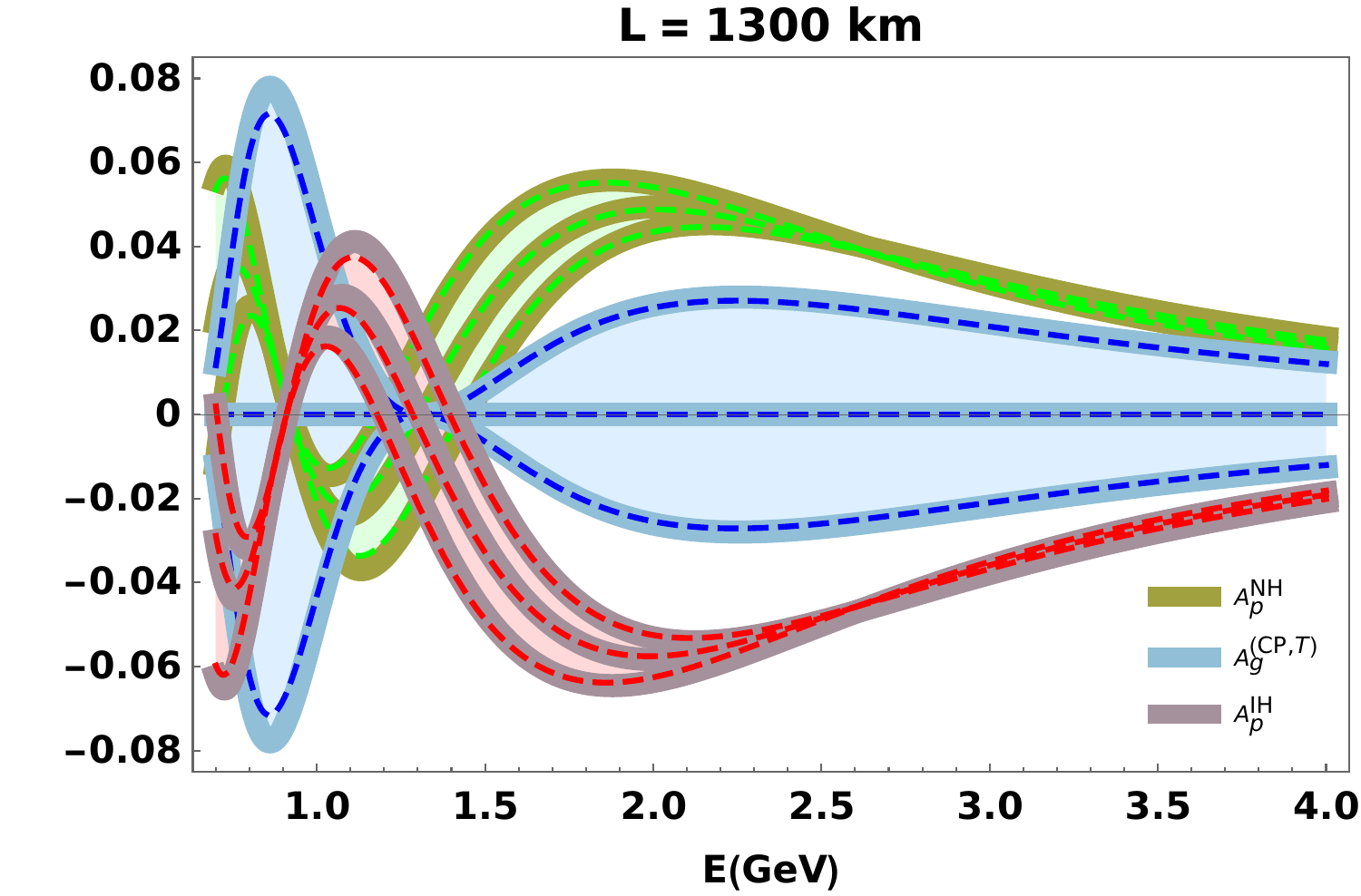}%
	\includegraphics[width=0.5\textwidth]{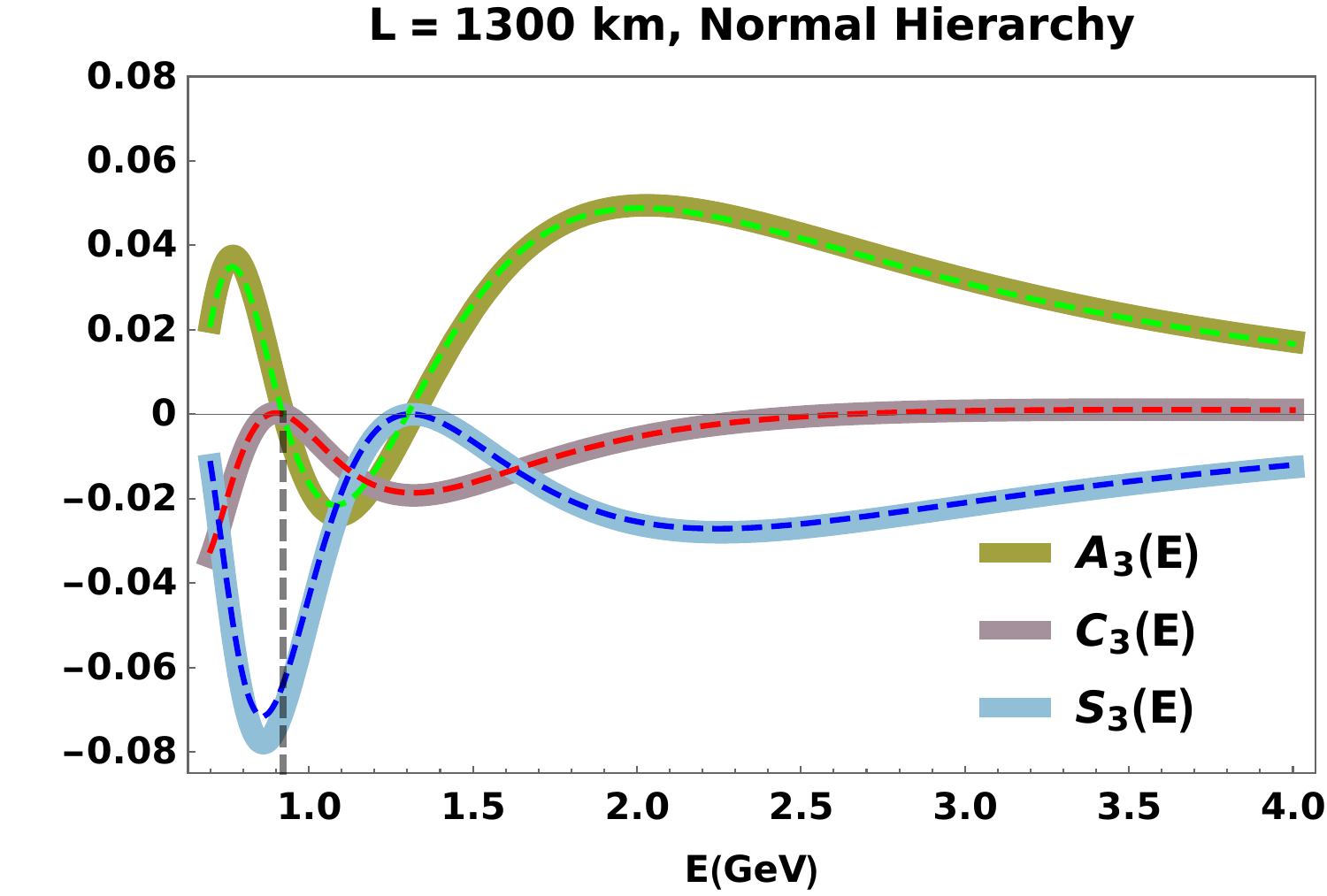}
	\caption{Left panel: The two components $A_p^{(CP, CPT)}(E)$ for normal 
		- green - and inverted  - red - hierarchies, as well                       
		as $A_g^{(CP, T)}(E)$ - blue -, of the CPV-asymmetry.
		Right panel: The separate $A_3(E)$, $C_3(E)$ and $S_3(E)$ terms,
		the black vertical dashed line depicts the location of the magic energy.
		Solid (dashed) lines portrait the approximate (numerical exact) equations.
		Colored bands depict the whole range of values of $\delta$.
	}\label{fig3}
\end{figure}

In the left panel of Fig. \ref{fig3} we give the separate results for $A_p^{(CP, CPT)}(E)$ for normal
- green - and inverted - red - hierarchies, with the bands covering all values of $\delta$, as well as $A_g^{(CP, T)}(E)$, with the blue band covering all values of $\delta$. They correspond to a baseline $L = 1300$ Km. of the DUNE experiment, showing larger effects of the matter potential. The dominance of $A_3(E)$ over the other components for energies above the first oscillation node is demonstrated in the right panel of Fig. \ref{fig3} with a separate plot of $A_3(E)$, $C_3(E)$ and $S_3(E)$.

We conclude that the dominance of $A_p^{(CP, CPT)}(E)$ over $A_g^{(CP, T)}(E)$ around
the first oscillation peak, with the magnitude of $A_3$ much higher than $C_3$, implies that {\hi{the sign of the observed total CPV-asymmetry $A^{(CP)}$ dictates the neutrino mass hierarchy independent of $\delta$}}. This conclusion holds for the expected value of the potential phase-shift $A$ present in $A_p^{(CP, CPT)}$ and absent in $A_g^{(CP, T)}$. Staying at energies around the first oscillation peak only, this conclusion is weakened without assuming the value of $A$:  the inequalities
$[ A^{(CP)} - |S_3| ] < Ap < [ A^{(CP)} + |S_3| ]$  
keep the conclusion that the sign of 
$A^{(CP)}$
fixes the hierarchy if 
$|A^{(CP)}| > |S_3|$.

For the ideal aim of this paper, an explicit measurement of the magnitude of the phase-shift $A$, needing the experimental separation of $A_p$ and $A_g$, can be made in the same experiment by {\hi{energy-dependence}} moving to energies around the second oscillation peak. The different energy dependence of the two components leads to the {\hi{magic energy}}
\autocite{Bernabeu:2018twl,Bernabeu:2018use} where the component $A_p^{(CP, CPT)}$ due to the potential vanishes for all $\delta$ values. The observed total CPV-asymmetry $A^{(CP)}$ around the magic energy at $E = 0.92$ GeV measures $A_g^{(CP, T)}$ and thus $\delta$. 
The magic energy in the CPV-asymmetry portrayed in this section  % (3) 
is the same than that discussed for the other asymmetries in the previous sections.

\section{Conclusion}\label{sec13}

To summarize, this research demonstrates that a test of the Aharonov-Bohm effect for a constant scalar potential, independent of any force-field, can be performed with the appropriate interpretation of the CPV asymmetry in the golden transition for the DUNE experiment on neutrino oscillations. The phase-shift due to the potential appears from the different constant potential of the e-flavour neutrinos with respect to the other flavours, not from a phase difference in space. The observable effect comes from flavour interferometry, not from spatial interference. Its measurement would prove without any loopholes the local effect of the potential in quantum mechanics.  {\hi{The potential is a physical property as it is the phase of probability amplitudes.}}

We have discussed how to signal and identify the phase-shift due
to the potential by means of the CPT-odd component of the discrete
asymmetries. The best case in existing experimental facilities corresponds to this component in the measurement of the CPV asymmetry in the $\nu_\mu \rightarrow \nu_e$ transition at the baseline of the DUNE experiment. The different energy dependence of the disentangled CPT-odd and CPT-even components of the CPV-asymmetry allows, by the comparison of the measurements in the first and second oscillation peaks, to obtain an
impressive observation of three fundamental open questions in physics:
CP violation in the lepton sector, the physical significance of the potential in quantum mechanics and the neutrino mass hierarchy.

\section*{Acknowledgements}

The authors would like to thank Antonio Di Domenico and Alejandro
Segarra for their comments and suggestions at different stages of this work.
C.E. would like to thank the Department of Theoretical Physics at the University of Valencia
for hospitality during part of this project.
This research has been funded by the Grants
CIPROM/2021/054 (Prometeo, Generalitat Valenciana), PID2023-151418NB-I00 (MCIU/AEI/10.13039/501100011033/ FEDER, UE) and CEX2023-001292-S
(SO, MCIU/AEI).
C.E. acknowledges the support of CONAHCYT (M\'exico) C\'atedra 341 
and in part by Mexican grants CONAHCYT CBF2023-2024-548 and UNAM PAPIIT IN111224.

\printbibliography

\end{document}